\documentclass[aps,prl,groupedaddress,twocolumn]{revtex4-1}
\usepackage{makeidx}
\usepackage{graphicx}
\usepackage{epsf}
\usepackage{amssymb,latexsym,amsmath}     

\begin{document}

\title{Computational toolbox for optical tweezers in geometrical optics}

\author{Agnese Callegari}
\email{agnese.callegari@fen.bilkent.edu.tr}
\affiliation{Soft Matter Lab, Department of Physics, Bilkent University, Cankaya, Ankara 06800, Turkey.}

\author{Mite Mijalkov}
\affiliation{Soft Matter Lab, Department of Physics, Bilkent University, Cankaya, Ankara 06800, Turkey.}

\author{A. Burak G\"ok\"oz}
\affiliation{Soft Matter Lab, Department of Physics, Bilkent University, Cankaya, Ankara 06800, Turkey.}

\author{Giovanni Volpe}
\email{giovanni.volpe@fen.bilkent.edu.tr}
\homepage{http://www.softmatter.bilkent.edu.tr}
\affiliation{Soft Matter Lab, Department of Physics, Bilkent University, Cankaya, Ankara 06800, Turkey.}

\date{\today}

\begin{abstract}
Optical tweezers have found widespread application in many fields, from physics to biology. Here, we explain in detail how optical forces and torques can be described within the geometrical optics approximation and we show that this approximation provides reliable results in agreement with experiments for particles whose characteristic dimensions are larger than the wavelength of the trapping light. Furthermore, we provide an object-oriented software package  implemented in MatLab for the calculation of optical forces and torques in the geometrical optics regime: \texttt{OTGO - Optical Tweezers in Geometrical Optics}. We provide all source codes for \texttt{OTGO} as well as the documentation and code examples -- e.g., standard optical tweezers, optical tweezers with elongated particle, windmill effect, Kramers transitions between two optical traps -- necessary to enable users to effectively employ it in their research and teaching.
\end{abstract}

\maketitle

\section{Introduction}

Optical tweezers are tightly focused laser beams capable of holding and manipulating microscopic particles in three dimensions. Since their invention in 1986 \cite{Ashkin1986}, optical tweezers have been increasing and consolidating their importance in several fields, from physics to biology \cite{Ashkin1997,Neuman2008,Dholakia2011,Padgett2011,Juan2011,Marago2013}. In the last fifteen years, thanks to the development of relatively simple and cheap setups, optical tweezers have also started to be employed in undergraduate and graduate laboratories as a tool to introduce students to advanced experimental techniques \cite{Smith1999,Bechhoefer2002,Appleyard2007,Volpe2013}.

Part of the reason for the success of optical tweezers lies in that the forces they can exert -- from tens of piconewtons down to tens of femtonewtons -- are just in the correct order of magnitude for a gentle but effective manipulation of colloidal particles and biological samples \cite{Ashkin1997,Neuman2008,Dholakia2011,Padgett2011,Juan2011,Marago2013}. An accurate mathematical description of these forces requires the use of electromagnetic theory in order to model the interaction between an incoming electromagnetic wave and a microscopic particle \cite{Borghese2003,Nieminen2007,Simpson2011}. However, this can be a daunting task. Therefore, it comes handy that simpler theoretical approaches have been shown to deliver accurate results in the limits where the particle characteristic dimensions are much smaller or much larger than the wavelength of the trapping light \cite{Neto2000}, which is typically between $532\,{\rm nm}$ and $1064\,{\rm nm}$ for optical tweezing applications. For particles much smaller than the wavelength, one can make use of the dipole approximation, which has already been extensively described and employed to describe the trapping of nanoparticles \cite{Marago2013}. For particles much larger than the wavelength, such as cells and large colloidal particles, whose size is typically significantly larger than one micrometer, one can make use of geometrical optics for the calculation of optical forces \cite{Ashkin1992}. This approach has been successfully employed, for example, to describe optical forces acting on cells \cite{Chang2006}, the deformation of microscopic bubbles in a optical field \cite{Skelton2012}, the optical lift effect \cite{Swartzlander2010} and the emerging of negative optical forces \cite{Kajorndejnukul2013}.

In this paper, we explain in detail how geometrical optics can be employed in order to study the optical forces and torques arising in an optical tweezers. We will first introduce how optical tweezers can be modeled in geometrical optics. Then, we will study in detail the forces associated to the scattering of a ray and of an optical beam by a spherical particle, distinguishing between scattering and gradient forces. Finally, we will explore some more complex situations, such as the arising of torque on non-spherical objects and the emergence of Kramers transitions between two optical tweezers. As an integral part of this article, we provide a complete MatLab software package -- \texttt{OTGO - Optical Tweezers in Geometrical Optics} -- to perform the calculation of optical forces and torques within the geometrical optics approach \cite{codes}.  \texttt{OTGO} is fully documented, accompanied by code examples and ready to be employed to explore more complex situations, both in research and in teaching. In fact,  we have implemented \texttt{OTGO} using an object-oriented approach so that it can be easily extended and adapted to the specific needs of users; for example, it is possible to create more complex optically trappable particles by extending the objects provided for spherical, cylindrical and ellipsoidal particles. In particular, we have used \texttt{OTGO} to obtain all the results presented in this article.


\section{Geometrical optics model of optical tweezers}

A schematic of a typical optical tweezers is shown in Fig.~\ref{fig:OTsetup}(a) and in supplementary movies 1, 2 and 3 \cite{codes}. A laser beam is focused by a high-NA objective (${\rm O_1}$) in order to create a high-intensity focal spot where a microscopic particle can be trapped. Typically, the particle is a dielectric sphere with refractive index  $n_{\rm p}$ immersed in a liquid medium with refractive index $n_{\rm m}$. The scattering of the focused beam on the particle generates some optical restoring forces that keep the particle near the focus. The sum of the incoming and scattered electromagnetic fields can be collected by a second objective (${\rm O_2}$) and projected onto a screen placed in the back-focal plane. The position of the optically trapped particle can be detected by using the image on the screen \cite{Ghislain1994}, as shown in Figs.~\ref{fig:OTsetup}(b) and \ref{fig:OTsetup}(c). Note that Fig.~\ref{fig:OTsetup} is not to scale by a factor $\sim 100$ because, in an actual setup, the objective focal length is $\sim 170\,{\rm \mu m}$ and the particle size is typically $\sim 2\,{\rm \mu m}$.

In the geometrical optics approach \cite{Ashkin1992}, the incoming laser beam, whose intensity profile is shown on the left of Figs.~\ref{fig:OTsetup}(a)-\ref{fig:OTsetup}(c), is decomposed into a set of optical rays, which are then focused by the objective ${\rm O_1}$. As the rays reach the particle, they get partially reflected and partially transmitted. The direction of the reflected and transmitted rays are different from those of the incoming rays. This change of direction entails a change of momentum and, because of the action-reaction law, a force acting on the sphere. As we will see, if $n_{\rm p}>n_{\rm m}$, these optical forces tend to pull the sphere towards the equilibrium position near the focal point. As the scattered rays reach the objective ${\rm O_2}$, they are collected and projected onto the back-focal plane.

%
%
\begin{figure}[h!]  
\centering
    \includegraphics[width=7cm]{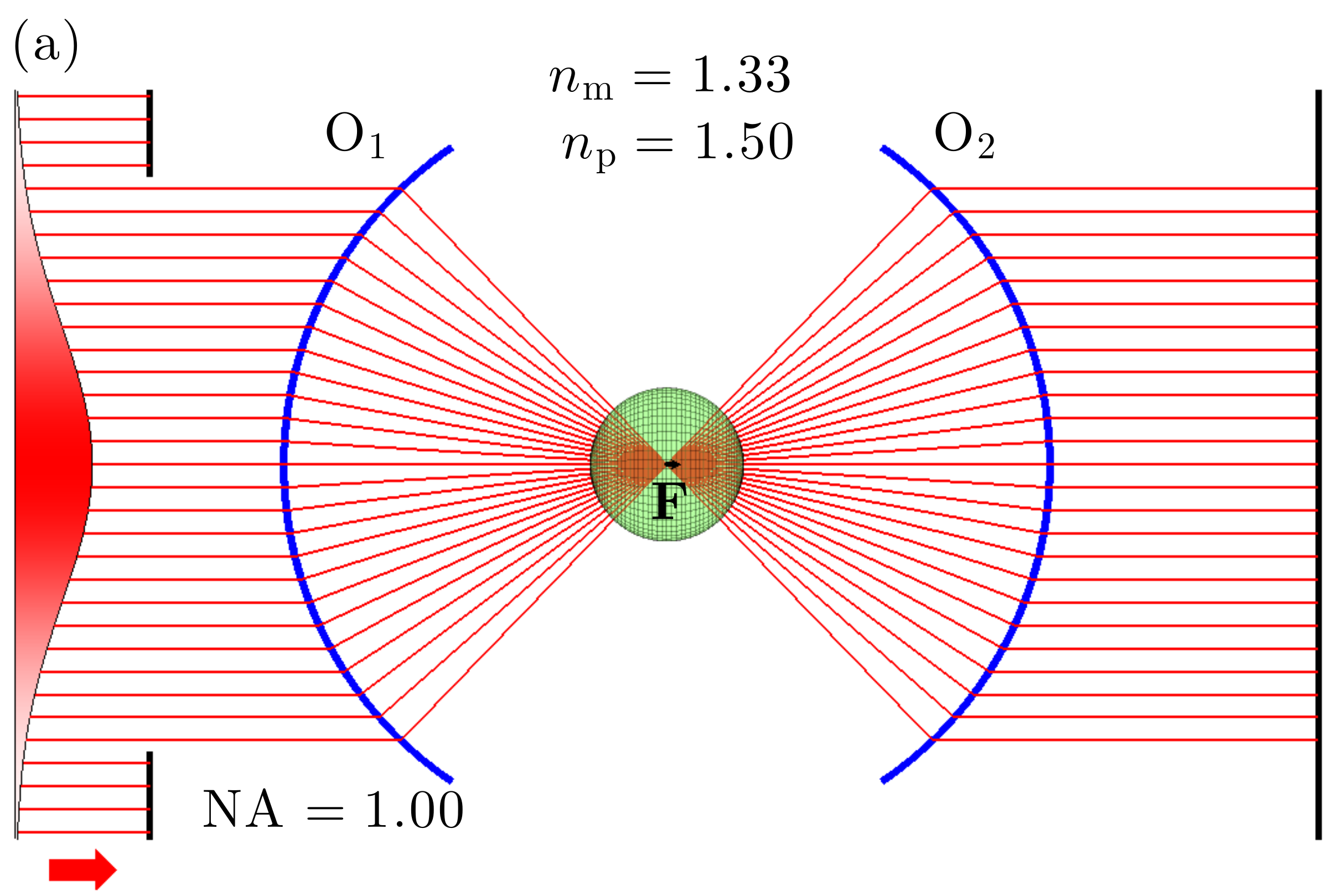}\\
    \includegraphics[width=7cm]{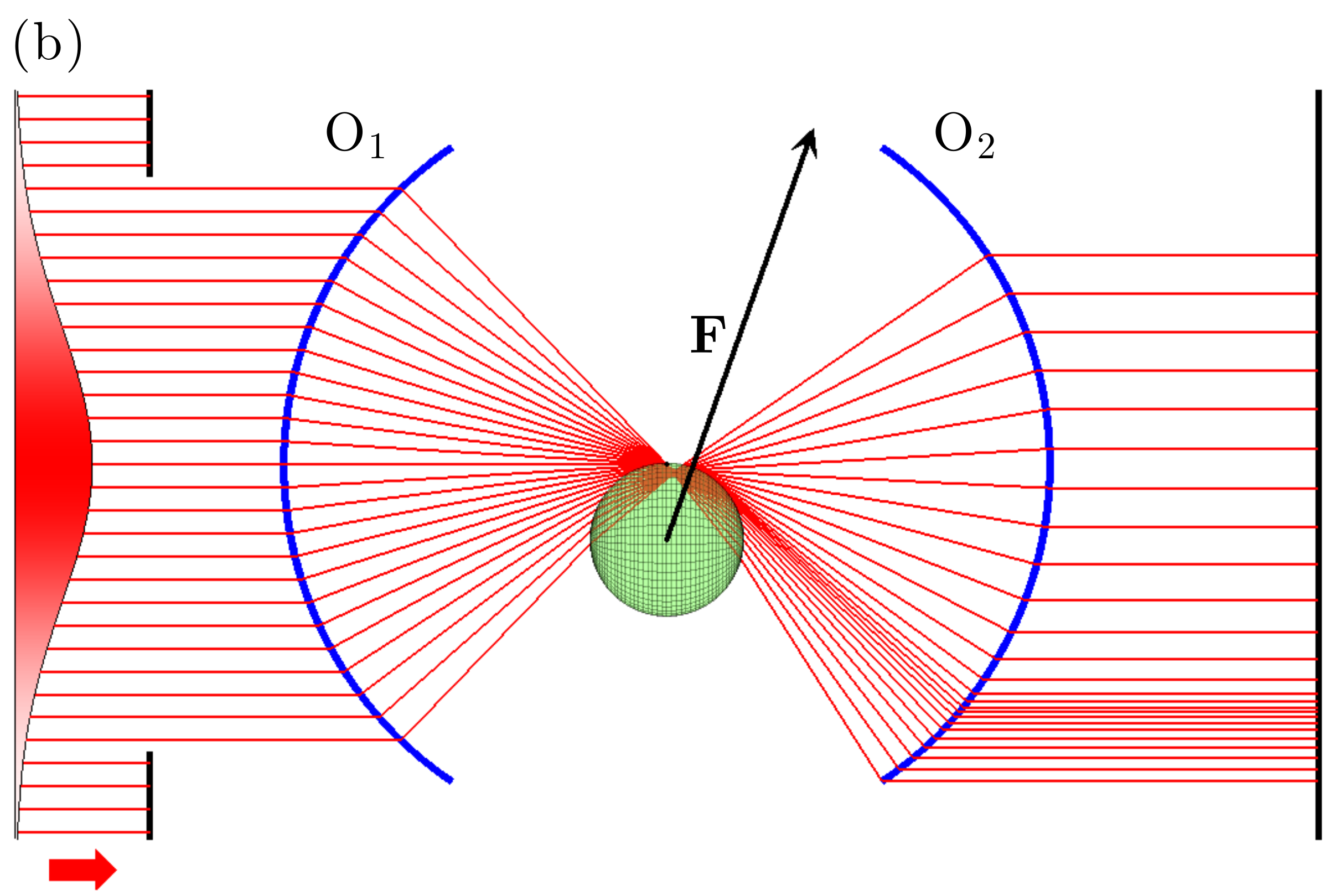}\\
    \includegraphics[width=7cm]{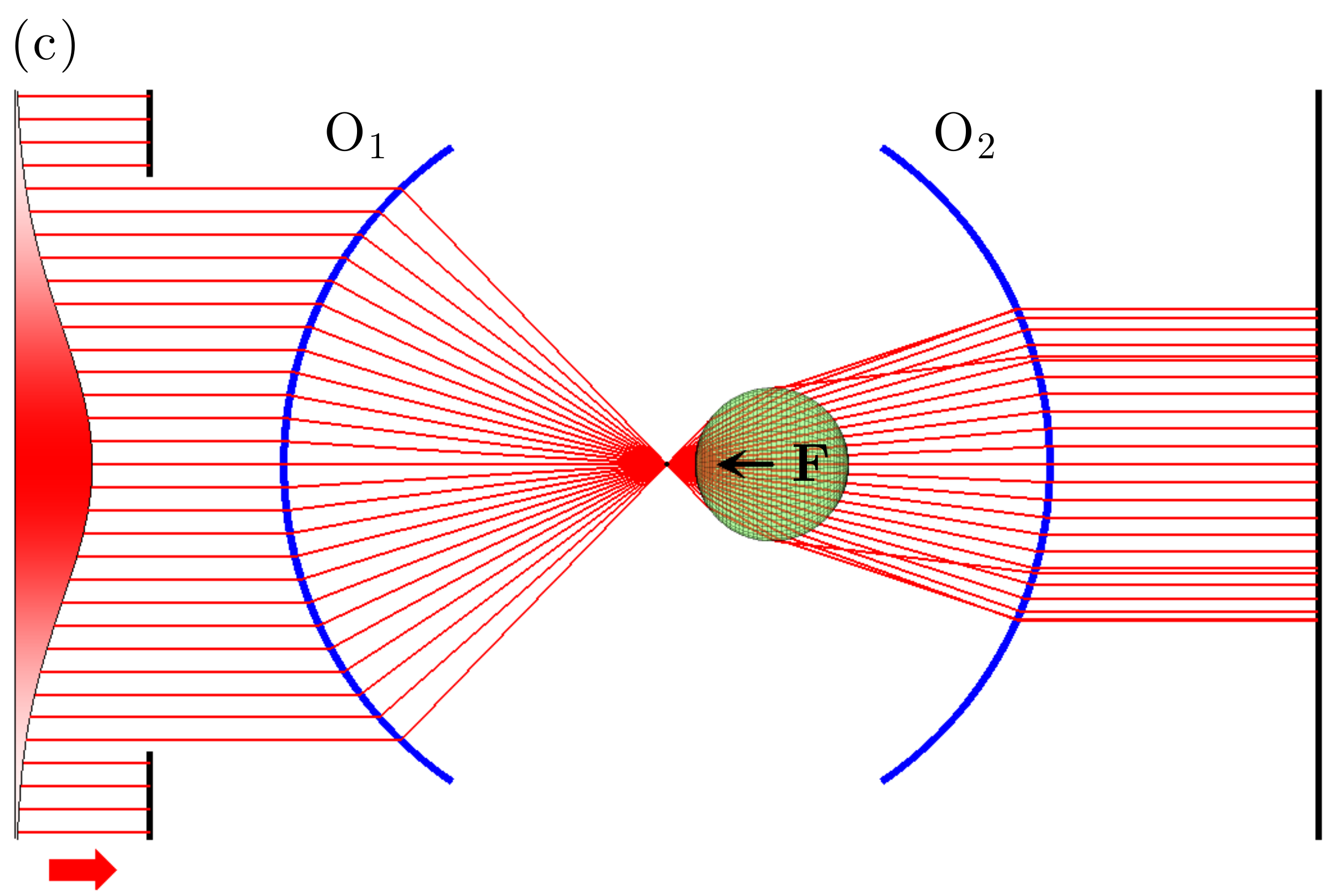}
\caption{(color online). Schematic of an optical tweezers setup (not to scale). A Gaussian laser beam, whose intensity is shown on the left, is divided into a set of optical rays (lines). The rays that can cross an aperture stop, whose radius is equal to the beam waist in this case, are then focused by an objective (${\rm O_1}$, ${\rm NA=1.00}$ in water). Near the focal point, a dielectric spherical particle (refractive index $n_{\rm p}=1.50$) immersed in a fluid (refractive index $n_{\rm m}=1.33$) scatters the rays (for clarity, the reflection of the incoming beam and the internally scattered rays are omitted) and, therefore, experiences a restoring recoil optical force ${\bf F}$ (black arrow). The scattered rays are collected by a second objective (${\rm O_2}$) and projected onto a screen placed in the back-focal plane. The position of the particle can be tracked by monitoring the back-focal plane image, which sensitively depends on the position of the particle, e.g., (a) at the focal point, (b) displaced in the transverse plane and (c) displaced along the longitudinal direction. The distance between the objectives and the size of the particle are not to scale by a factor $\sim 100$. See also supplementary movies 1, 2 and 3 \cite{codes}.}
\label{fig:OTsetup}
\end{figure}


\section{Forces by a ray on a planar surface}

The energy flux transported by a monochromatic electromagnetic field, such as the one of a laser beam, is given by its Poynting vector
\begin{equation}
   {\bf S} = \frac{1}{2 \mu}{\rm Re} \left\{ {\bf E} \times  {\bf B}^{*} \right\},
\end{equation}
where ${\bf E}$ and ${\bf B}$ are the complex electric and magnetic fields. In order to describe how this energy is transported, a series of rays can be associated with the electromagnetic field \cite{goodman1995}. These rays are lines perpendicular to the electromagnetic wavefronts and pointing in the direction of the electromagnetic energy flow. 

When a light ray impinges on a flat surface between two media with different refractive indices, it is partly reflected and partly transmitted. Given an incidence angle ${\theta_{\rm i}}$, i.e., the angle between the incoming ray ${\bf r}_{\rm i}$ and the normal ${\bf n}$ to the surface at the incidence point, the reflection angle ${\theta_{\rm r}}$ is given by the reflection law
\begin{equation} \label{eqn:refl}
  \theta_{\rm r} = \theta_{\rm i}
\end{equation}
and the transmission angle ${\theta_{\rm t}}$ is given by Snell's law
\begin{equation} \label{eqn:Snell}
\theta_{\rm t} =  {\rm asin} \left( \frac{n_{\rm i}}{n_{\rm t}} \sin{ \theta_{\rm i} } \right),
\end{equation}
where $n_{\rm i}$ is the refractive index of the medium of the incident ray  ${\bf r}_{\rm r}$ and $n_{\rm t}$ is the one of the medium of the transmitted ray  ${\bf r}_{\rm t}$. Both ${\bf r}_{\rm r}$ and ${\bf r}_{\rm t}$ lie in the plane of incidence, i.e., the plane that contains ${\bf r}_{\rm i}$ and ${\bf n}$.
Because of energy conservation, the power $P_{\rm i}$ of  ${\bf r}_{\rm i}$ must be equal
to the sum of the power $P_{\rm r}$ of  ${\bf r}_{\rm r}$ and the power $P_{\rm t}$ of  ${\bf r}_{\rm t}$, i.e.,
\begin{equation} \label{eqn:PowerCons}
P_{\rm i} = P_{\rm r} + P_{\rm t}.
\end{equation}
How the power is split can be calculated by using Maxwell's equations with the appropriate boundary conditions \cite{someda2006electromagnetic}. The result is expressed by Fresnel's equations and depends on the polarization of the incoming ray, as we must distinguish the case when the electric field of the ray oscillates in the plane of incidence ($p$-polarization) from the one when it oscillates in a plane perpendicular to the plane of incidence ($s$-polarization). The Fresnel's  reflection and transmission coefficients for  $p$-polarized light are
\begin{equation} \label{eqn:FresnelReflP}
R_{\rm p} = \left|    \frac{  n_{\rm i} \cos{ \theta_{\rm t}} -  n_{\rm t} \cos{ \theta_{\rm i}}   }{n_{\rm i} \cos{ \theta_{\rm t}} +  n_{\rm t} \cos{ \theta_{\rm i}}}    \right|^{2},
\end{equation}
\begin{equation} \label{eqn:FresnelTransP}
T_{\rm p} =  \frac{ 4 n_{\rm i}  n_{\rm t} \cos{ \theta_{\rm i} } \cos{ \theta_{\rm t} }   }{\left|  n_{\rm i} \cos{ \theta_{\rm t} } +  n_{\rm t} \cos{ \theta_{\rm i} }\right|^{2}}
\end{equation}
and for $s$-polarized light
\begin{equation} \label{eqn:FresnelReflS}
R_{\rm s} = \left|    \frac{  n_{\rm i} \cos{ \theta_{\rm i} } -  n_{\rm t} \cos{ \theta_{\rm t} }   }{n_{\rm i} \cos{ \theta_{\rm i} } +  n_{\rm t} \cos{ \theta_{\rm t} }}    \right|^{2},
\end{equation}
\begin{equation} \label{eqn:FresnelTransS}
T_{\rm s} =  \frac{ 4 n_{\rm i}  n_{\rm t} \cos{ \theta_{\rm i} } \cos{ \theta_{\rm t} }   }{\left|  n_{\rm i} \cos{ \theta_{\rm i} } +  n_{\rm t} \cos{ \theta_{\rm t} }\right|^{2}}.
\end{equation}
For unpolarized and circularly polarized light, one can use the average of the previous coefficients, i.e.,
\begin{equation} \label{eqn:FresnelU}
R = \frac{R_{\rm p} + R_{\rm s}}{2},
\end{equation}
\begin{equation} \label{eqn:FresnelU}
T = \frac{T_{\rm p} + T_{\rm s}}{2}.
\end{equation}

The recoil optical forces are equal and opposite to the rate of change of linear momentum of the light. Since for a ray of power $P$ in a medium of refractive index $n$, the momentum flux is $nP/c$, where $c$ is the speed of light in vacuum, the optical force is \cite{Ashkin1992}
\begin{equation}\label{eq:FonSurface}
{\bf F}_{\rm pl} = \frac{n_{\rm i} P_{\rm i}}{c} \hat{\bf u}_{\rm i} - \frac{n_{\rm i} P_{\rm r}}{c}\hat{\bf u}_{\rm r} - \frac{n_{\rm t} P_{\rm t}}{c} \hat{\bf u}_{\rm t},
\end{equation}
where $\hat{\bf u}_{\rm i}$ is the unit vector of ${\bf r}_{\rm i}$, $\hat{\bf u}_{\rm r}$ is the unit vector of ${\bf r}_{\rm r}$ and $\hat{\bf u}_{\rm t}$ is the unit vector of ${\bf r}_{\rm t}$. We must note that the definition of the momentum of light in a medium is a thorny issue, which is often referred to as the {\em Abraham-Minkowski dilemma} after the works of Hermann Minkowski \cite{Minkowski1908} and Max Abraham \cite{Abraham1909}. This issue is discussed in detail, e.g., in Refs.~\cite{Pfeifer2007,Barnett2010}. Since  most results in optical trapping and manipulation do not depend qualitatively on the momentum definition, in this work we employ the Minkowski momentum definition  which in fact is the most often employed in optical tweezers studies \cite{Ashkin1992,Pfeifer2009}. However,
we remark that all results can be easily adapted to the Abraham momentum definition by changing the definition of the force in Eq.~(\ref{eq:FonSurface}) \cite{Pfeifer2007}.

\section{Forces by a ray on a sphere}

%
%
\begin{figure}[t!]
\centering
    \includegraphics[width=8.5cm]{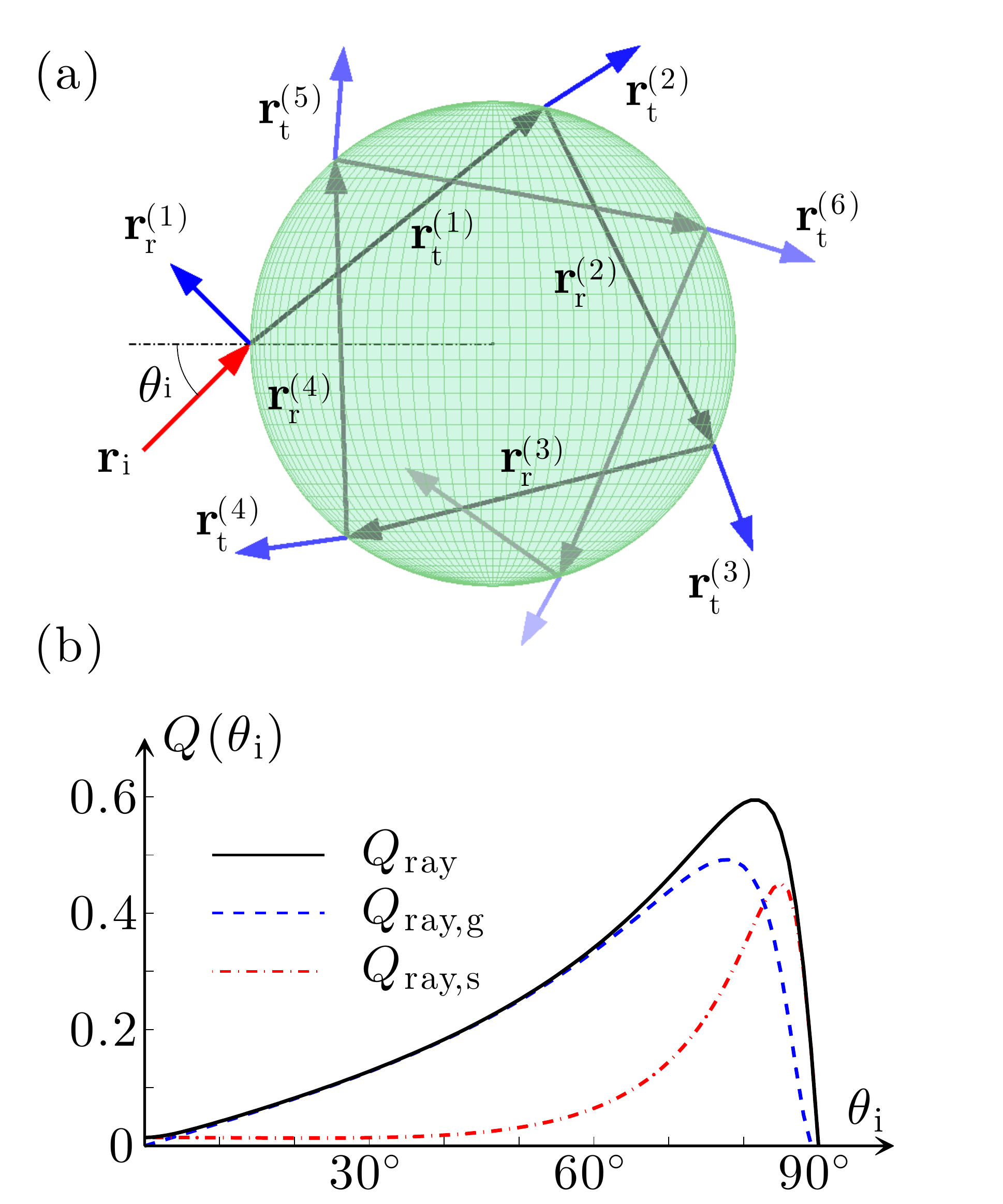}
\caption{(color online). (a) Scattering of a ray impinging on a sphere. The incident ray ${\bf r}_{\rm i}$ impinges on a glass spherical particle ($n_{\rm p} = 1.50$) immersed in water ($n_{\rm m} = 1.33$). The reflected (${\bf r}_{\rm r}^{(j)}$) and transmitted (${\bf r}_{\rm t}^{(j)}$) rays for the first seven scattering events are represented. Because of the spherical symmetry of the particle, all rays lie in the plane of incidence. See also supplementary movies 4 and 5 \cite{codes}.
(b) Corresponding trapping efficiencies as a function of the incidence angle $\theta_{\rm i}$. See also supplementary movies 6 and 7 \cite{codes}.}
\label{fig:RayOnSphere}
\end{figure}

We now consider a ray ${\bf r}_{\rm i}$ of power $P_{\rm i}$ impinging from a medium with refractive index $n_{\rm m}$ on a dielectric sphere with refractive index $n_{\rm p}$ at an incidence angle $\theta_{\rm i}$, as shown in Fig.~\ref{fig:RayOnSphere}(a) and in supplementary movies 4 and 5 \cite{codes}. As soon as  ${\bf r}_{\rm i}$ hits the sphere, a small amount of its power, $P_{\rm r}^{(1)}$, is diverted into the reflected ray ${\bf r}_{\rm r}^{(1)}$, while most power, $P_{\rm t}^{(1)}$, goes into the transmitted ray ${\bf r}_{\rm t}^{(1)}$. The ray ${\bf r}_{\rm t}^{(1)}$ crosses the sphere until it reaches the opposite surface, where again a large portion of its power, $P_{\rm t}^{(2)}$, is transmitted outside the sphere into the ray ${\bf r}_{\rm t}^{(2)}$, while a small amount of its power, $P_{\rm r}^{(2)}$, is reflected inside the sphere into the ray ${\bf r}_{\rm r}^{(2)}$. The ray ${\bf r}_{\rm r}^{(2)}$ undergoes another scattering event as soon as it reaches the sphere boundary, and the process continues until all light has escaped from the sphere. The force $ {\bf F}_{\rm ray}$ produced on the sphere by this series of scattering events can be calculated by using repeatedly Eq.~(\ref{eq:FonSurface}), i.e.,
\begin{equation}\label{eq:FonParticle}
 {\bf F}_{\rm ray} = \frac{n_{\rm m} P_{\rm i}}{c} \hat{\bf u}_{\rm i} - \frac{n_{\rm m} P_{\rm r}^{(1)}}{c}\hat{\bf u}_{\rm r}^{(1)} -  \sum_{j=2}^{\infty} \frac{n_{\rm m} P_{\rm t}^{(j)}}{c} \hat{\bf u}_{\rm t}^{(j)},
\end{equation}
where $\hat{\bf u}_{\rm i}$, $\hat{\bf u}_{\rm r}^{(1)}$ and $\hat{\bf u}_{\rm t}^{(j)}$ are the unit vectors of the incident ray, the first reflected ray and the $j$-th transmitted ray, respectively. We note that the dependence of Eq.~(\ref{eq:FonParticle}) on $n_{\rm p}$ is hidden in the dependence of the quantities $P_{\rm r}^{(1)}$ and $P_{\rm t}^{(j)}$ on the Fresnel's coefficients [Eqs.~(\ref{eqn:FresnelReflS}), (\ref{eqn:FresnelTransS}), (\ref{eqn:FresnelReflP}) and (\ref{eqn:FresnelTransP})]. Furthermore, we can notice that the absolute value of the force does not depend on the dimension of the particle.

Since all the reflected and transmitted rays are contained in the plane of incidence, as can be seen in Fig.~\ref{fig:RayOnSphere}(a) and in the supplementary movies 4 and 5 \cite{codes}, also the force ${\bf F}_{\rm ray}$ in Eq.~(\ref{eq:FonParticle}) has components only within the incidence plane. We can, therefore, split ${\bf F}_{\rm ray}$ into a component along the direction of the incoming ray, i.e., the {\em scattering force} ${\bf F}_{\rm ray, s} = ({\bf F}_{\rm ray} \cdot \hat{\bf u}_{\rm i}) \, \hat{\bf u}_{\rm i} = F_{\rm ray,s} \, \hat{\bf u}_{\rm i}$, and a component perpendicular to the direction of the incoming ray, i.e., the {\em gradient force} ${\bf F}_{\rm ray, g} = {\bf F}_{\rm ray} - ({\bf F}_{\rm ray} \cdot \hat{\bf u}_{\rm i}) \, \hat{\bf u}_{\rm i} = F_{\rm ray, g} \, \hat{\bf u}_{\perp}$:
\begin{equation}\label{eq:DefScatGradForce}
{\bf F}_{\rm ray} = {\bf F}_{\rm ray, s} + {\bf F}_{\rm ray, g} = F_{\rm ray, s} \, \hat{\bf u}_{\rm i} + F_{\rm ray, g} \, \hat{\bf u}_{\perp},
\end{equation}
where $\hat{\bf u}_{\perp}$ is the unit vector perpendicular to $\hat{\bf u}_{\rm i}$ and contained in the incidence plane. Interestingly, the gradient force is a conservative force, while the scattering force is nonconservative. If $n_{\rm p}>n_{\rm m}$, the particle is attracted towards the ray [supplementary movie 4 \cite{codes}], while, if $n_{\rm p}<n_{\rm m}$, the particle is pushed away from the ray [supplementary movie 5 \cite{codes}].

In order to quantify the effectiveness of the transfer of momentum from the ray to the particle, we can introduce the \emph{trapping efficiency}, i.e., the ratio between the modulus of the optical force and the momentum per second of the incoming ray in a medium with refraction index $n_{\rm i}$. The trapping efficiency is bound to lie between $0$, corresponding to a ray that is not deflected, and $2$, corresponding to a ray that is reflected back on its path \cite{Ashkin1992}. For example, for a $1\,{\rm mW}$ ray, the maximum optical force is $7 \cdot 10^{-12}\,{\rm N}$, i.e., 7 piconewtons. Albeit small, this force is comparable to the forces that are relevant in the microscopic and nanoscopic world, e.g., the forces generated by molecular motors \cite{Rohrbach2005}, and gives us a first impression of the potential of optical manipulation. In particular, we can define the \emph{scattering trapping efficiency} 
\begin{equation}\label{eq:Qs}
Q _{\rm ray, s}= \frac{c}{n_{\rm i} P_{\rm i}} F_{\rm ray, s},
\end{equation}
the \emph{gradient trapping efficiency}
\begin{equation}\label{eq:Qg}
Q _{\rm ray, g}= \frac{c}{n_{\rm i} P_{\rm i}} F_{\rm ray, g}
\end{equation}
and the total scattering efficiency
\begin{equation}\label{eq:Q}
Q_{\rm ray}  = \sqrt{Q _{\rm ray, g}^2 + Q _{\rm ray, s}^2 }.
\end{equation}
Fig.~\ref{fig:RayOnSphere}(b) and supplementary movie 6 \cite{codes} show the trapping efficiencies as a function of $\theta_{\rm i}$ for a circularly polarized ray impinging on a glass sphere ($n_{\rm p} = 1.50$) immersed in water ($n_{\rm m} = 1.33$); supplementary movie 7 \cite{codes} shows the trapping efficiencies for a circularly polarized ray impinging on an air bubble  ($n_{\rm p} = 1.00$) immersed in water ($n_{\rm m} = 1.33$). In both cases, the major contribution to the total trapping efficiency is given by $Q_{\rm ray, g}$, while only for very large incidence angles $Q_{\rm ray, s}$ becomes appreciable. 


\section{Forces by a focused beam on a sphere}

%
%
\begin{figure}[b]  
\centering
	\includegraphics[width=8.5cm]{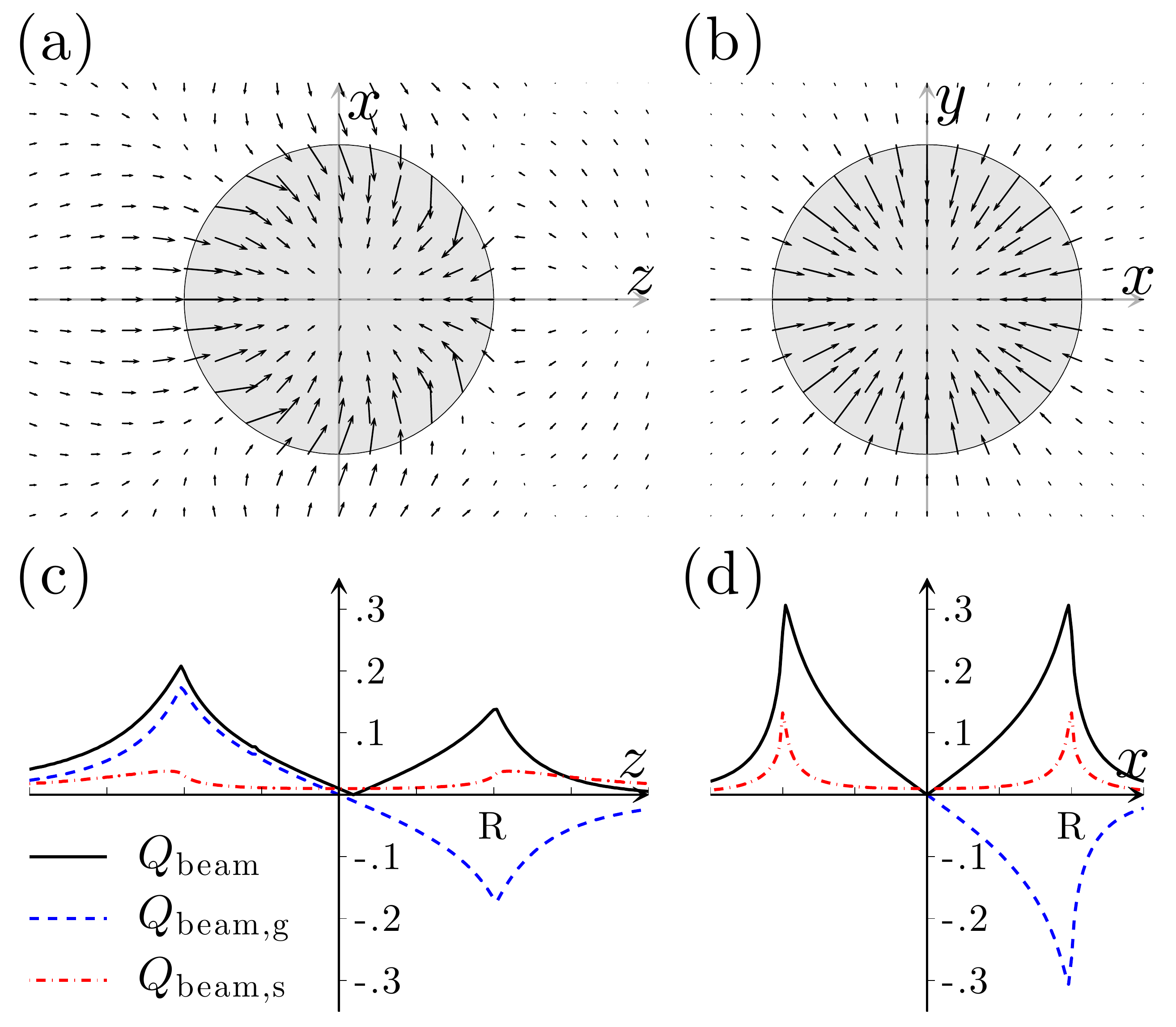}
\caption{(color online). Optical force fields in an optical tweezers. The arrows represent the direction and magnitude of the force exerted on a glass spherical particle ($n_{\rm p}=1.50$) in water ($n_{\rm m}=1.33$) illuminated by a highly focused Gaussian beam (${\rm NA}=1.30$, beam waist equal to the aperture stop radius) propagating along the $z$-direction as a function of the particle position (a) in the longitudinal ($zx$) plane and (b) in the transverse ($xy$) plane. The shaded area represents the dimension of the particle. The corresponding trapping efficiencies are shown in (c) and (d) for particle displacements along the $z$-axis and $x$-axis, respectively. Note that for displacements along the $z$-axis both the scattering and the gradient force are directed along $z$, while for displacements along the $x$-axis the gradient force is directed along $x$, but the scattering force is directed along $z$. }
\label{fig:ForceFieldandTrapEffSphe}
\end{figure}

%
%
\begin{figure}[t]  
\centering
\includegraphics[width=8.5cm]{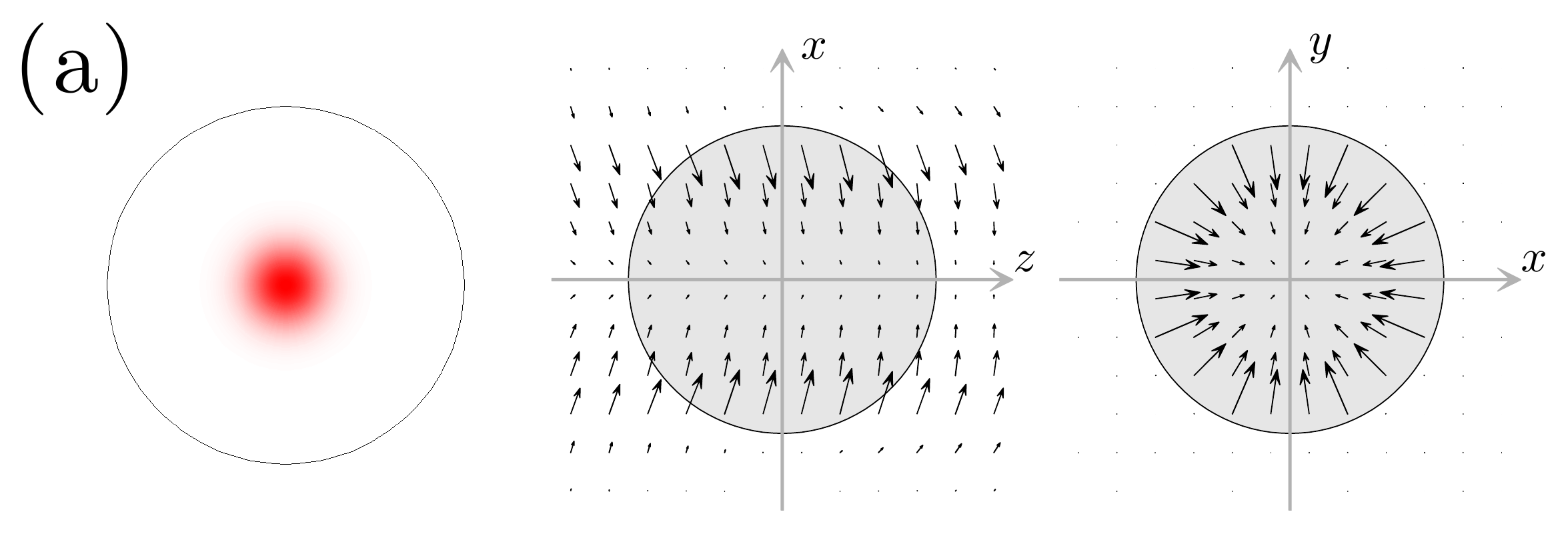}
\includegraphics[width=8.5cm]{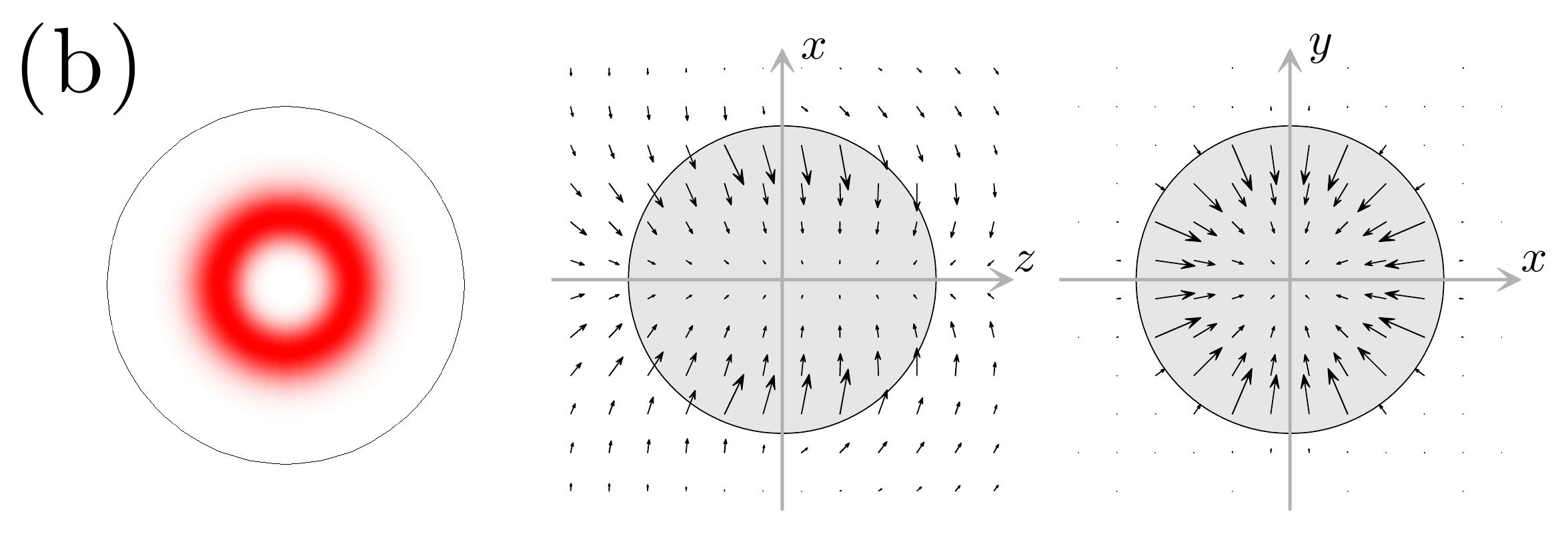}
\includegraphics[width=8.5cm]{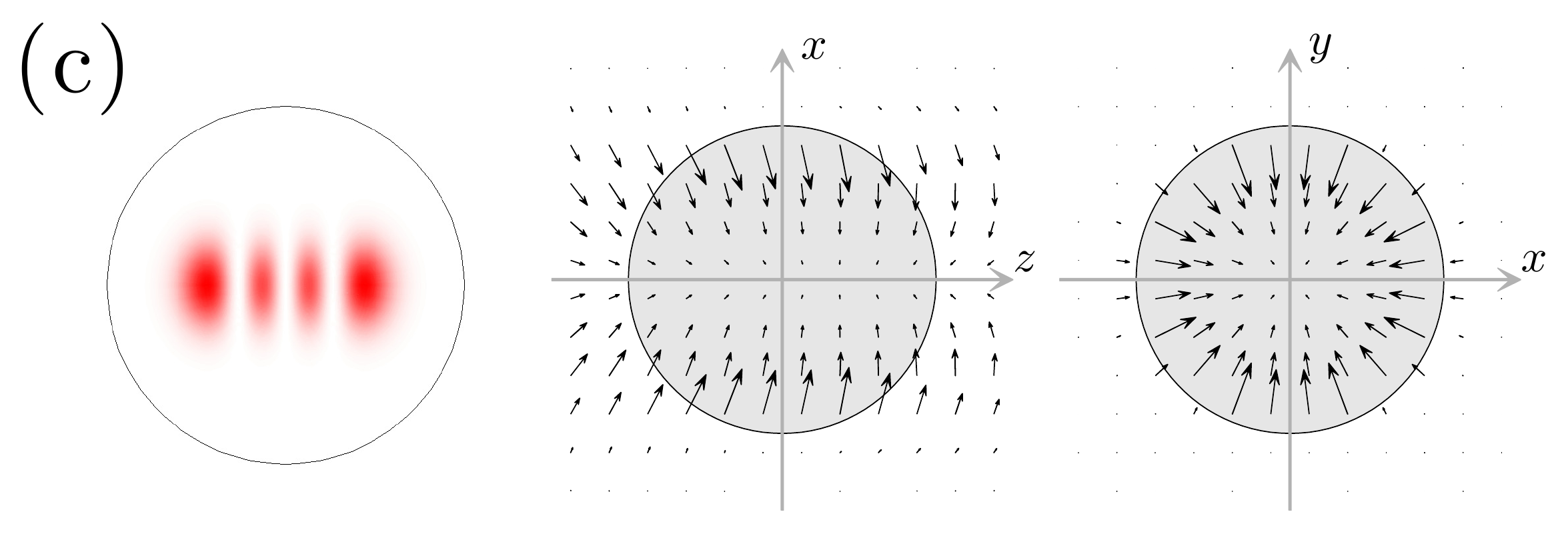}
\caption{(color online). Force fields in an optical tweezers generated using various kinds of beams. (a) A Gaussian beam with a waist much smaller than the radius of the aperture stop can trap a particle in the transverse plane but cannot confine the particle along the longitudinal direction.
(b) A Laguerre-Gaussian beam, or doughnut beam, improves the trapping along the longitudinal direction because of the presence of more power at large angles.
(c) A Hermite-Gaussian beam clearly breaks the cylindrical symmetry of the trap, as can be seen from the force field in the transverse plane.
In all cases, the force fields are calculated for a glass spherical particle ($n_{\rm p}=1.50$) in water ($n_{\rm m}=1.33$). The circle on the left corresponds to ${\rm NA}=1.30$.
}
\label{fig:ForceFieldandTrajDiffBeams}
\end{figure}

It is not possible to achieve a stable trapping using a single ray because the particle is permanently pushed by the scattering force in the direction of the incoming ray, as we have seen in Fig.~\ref{fig:RayOnSphere}(b). A possible approach to achieve a stable trap is to use a second counter-propagating light ray. In fact, such a configuration using two laser beams was amongst the first ones to be employed in order to trap and manipulate microscopic particles \cite{Ashkin1970} and a modern version has been obtained using the light emerging from two optical fibers facing each other \cite{Jess2006}. This approach works also if the two beams are not perfectly counter-propagating, but they are arranged with a sufficiently large angle. 

A more convenient alternative to using several counter-propagating light beams is to use a single highly-focused light beam. In fact, rays originating from diametrically opposite points of a high-NA focusing lens produce in practice a set of rays that converge at a very large angle, as can be seen in Fig.~\ref{fig:OTsetup}.

The most commonly employed laser beam is a Gaussian beam. Its intensity profile at the waist is given by 
\begin{equation}\label{eq:intbg}
I^{\rm G}(\rho) = I_{\rm 0} e^{ -\frac{\rho^2}{2 w_0^2} },
\end{equation} 
where $\rho$ is the radial coordinate, $w_0$ is the beam waist, $I_{\rm 0} = \frac{1}{2} c \varepsilon_0 n_{\rm m} E_{\rm 0}^{2}$ is the beam intensity at $\rho=0$, $\varepsilon_0$ is the dielectric permittivity of vacuum, and $E_0$ is the modulus of the electric field magnitude at $\rho=0$. Such a beam can be approximated by a set of rays parallel to the optical axis ($z$) each endowed with a power proportional to the local intensity of the beam. The resulting rays are then focused by an objective lens, which has the effect of bending the light rays towards the focal point, as shown in Fig.~\ref{fig:OTsetup} and supplementary movies 1, 2 and 3 \cite{codes}. Each one of these rays produces a force ${\bf F}_{\rm ray}^{(m)} $ on the sphere given by  Eq.~(\ref{eq:FonParticle}). The total optical force exerted by the focused beam on the sphere is then the sum of all the rays' contributions, i.e.,
\begin{equation}{\label{eq:FBeamonParticle}}
 {\bf F}_{\rm beam} = \sum_{m} {\bf F}_{\rm ray}^{(m)} 
\end{equation}
In Fig.~\ref{fig:ForceFieldandTrapEffSphe}(a) and \ref{fig:ForceFieldandTrapEffSphe}(b), the force fields in the longitudinal ($zx$) and transverse ($xy$) planes are represented as a function of the distance of the high-refractive index spherical particle ($n_{\rm p}=1.50$, $n_{\rm m}=1.33$) from the focal point, in the case of an objective with $\mathrm{NA}=1.30$ and a circularly polarized beam. 

To have a comparison of the force obtained from our simulations with the forces usually found in experiments, 
we can compare our prediction with the results in Ref. \cite{Rohrbach2005}. We take, for comparison, the measured trapping stiffnesses for a polystyrene sphere ($n_{\mathrm p}=1.57$) with a 1.66 $\mu{\mathrm m}$ diameter in a trap generated using a Gaussian beam with power $P=10\ \rm{mW}$ focused by an objective with $\mathrm{NA}=1.20$. Performing a calculation with the cited parameters,  we obtain a trapping stiffness along the longitudinal direction ($z$) equal to $k_{z}^{\texttt{OTGO}} = 6.45\ \mathrm{pN/}\mu{\mathrm m}$, and a trapping stiffness along the transversal direction ($x$) equal to $k_{x}^{\texttt{OTGO}} = 12.66\ \mathrm{pN/}\mu{\mathrm m}$, in reasonable agreement with the experimental values found in Ref. \cite{Rohrbach2005}, which are respectively $k_z^{\texttt{exp}}=3.85\ \mathrm{pN/}\mu{\mathrm m}$ and $k_x^{\texttt{exp}}=11.0\ \mathrm{pN/}\mu{\mathrm m}$.

The optical force field is cylindrically symmetric around the $z$-axis. The equilibrium position lies on $z$-axis, i.e., $(x,y)=(0,0)$, and is slightly displaced towards positive $z$ because of the presence of scattering forces, as is commonly observed in experiments \cite{Merenda2006}. In fact, a Brownian particle in an optical trap is in dynamic equilibrium with the thermal noise pushing it out of the trap and the optical forces driving it toward the center of the trap \cite{Volpe2013}, as can be seen from the Brownian motion that the particles experience in supplementary movies 1, 2 and 3 \cite{codes}. The maximum value of the force is achieved when the particle displacement is about equal to the particle radius $R$. We can notice again that, like in the case of a single ray, the value of the force does not depend on $R$; however, the trap stiffness, which is given by the force divided by the displacement is inversely proportional to $R$.

The force ${\bf F}_{\rm beam}$ in Eq.~(\ref{eq:intbg}) can be split into a scattering force ${\bf F}_{\rm beam, s}$, given by the sum of  the scattering forces of each ray, and a gradient force ${\bf F}_{\rm beam, g}$, given by the sum of the gradient forces of each ray \cite{Ashkin1992}, i.e., 
\begin{equation}
 {\bf F}_{\rm beam, s} = \sum_{m} {\bf F}_{\rm ray, s}^{(m)}
\end{equation}
and
\begin{equation}
{\bf F}_{\rm beam, g} = \sum_{m} {\bf F}_{\rm ray, g}^{(m)}.
\end{equation}
Like in the case of a single ray, while the gradient force ${\bf F}_{\rm beam, g} $ is conservative, the scattering force ${\bf F}_{\rm beam, s}$ can give rise to nonconservative effects, as has been shown in various experiments \cite{Merenda2006}. These nonconservative effects are, however, small \cite{Pesce2009} as can be seen from the small displacement along the $z$-axis of the equilibrium position.

It is now possible to define the scattering efficiencies for the focused beam as
\begin{equation}
Q_{\rm beam, s} =  \frac{c}{n_{\rm i} P_{\rm beam}} F_{\rm beam, s},
\end{equation}
\begin{equation}
Q_{\rm beam, g} = \frac{c}{n_{\rm i} P_{\rm beam}} F_{\rm beam, g}
\end{equation}
and
\begin{equation}
Q_{\rm beam} = \frac{c}{n_{\rm i} P_{\rm beam}} F_{\rm beam},
\end{equation}
where $P_{\rm beam}$ is the power of beam that contributes to the focal fields, i.e., after the aperture stop. The scattering coefficients are shown in Fig.~\ref{fig:ForceFieldandTrapEffSphe}(c) and \ref{fig:ForceFieldandTrapEffSphe}(d) for a sphere displaced along the longitudinal and transverse direction, respectively. If the sphere is on the $z$-axis, i.e., the propagation axis of the beam, both the scattering force and the gradient force act only along the $z$-direction because of symmetry. For displacements of the particle along the $x$-direction, the gradient force is along the $x$-direction and the scattering force along the $z$-direction. 

Other kinds of beams can also be used in optical trapping experiments. In particular, Laguerre-Gaussian and Hermite-Gaussian beams \cite{someda2006electromagnetic} have been widely exploited. An accurate description of these beams requires one to take into account their orbital angular momentum \cite{andrews2012angular}, which can have major effects on their trapping properties. However, the features related to the presence of spin angular momentum and of a non-uniform phase profile in the beam, which lead, e.g., to the presence of orbital angular momentum, cannot be accurately modeled within the geometrical optics approach. Nevertheless, some features connected to the different intensity distribution in Laguerre-Gaussian and Hermite-Gaussian beams can be explored, as shown in Fig.~\ref{fig:ForceFieldandTrajDiffBeams}.


\section{Further numerical experiments}
 
This section provides some guidelines and examples on how readers can use geometrical optics and \texttt{OTGO} to explore more complex situations both in the lab and in the classroom, going beyond the basic optical tweezers case of a microscopic sphere optically trapped in a highly focused laser beam.

%
%
\begin{figure}[ht!]  
\centering
\includegraphics[width=2.8cm]{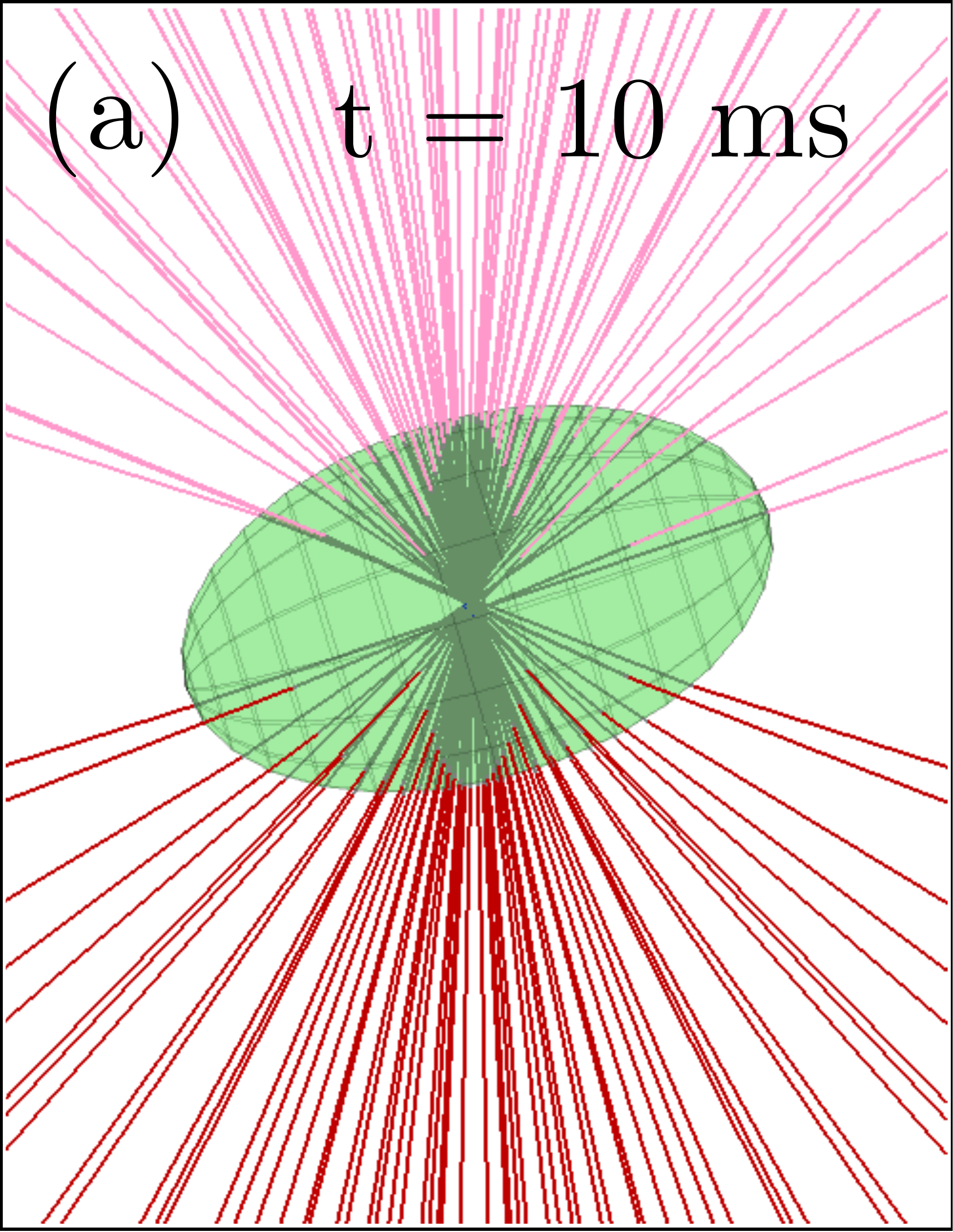}
\includegraphics[width=2.8cm]{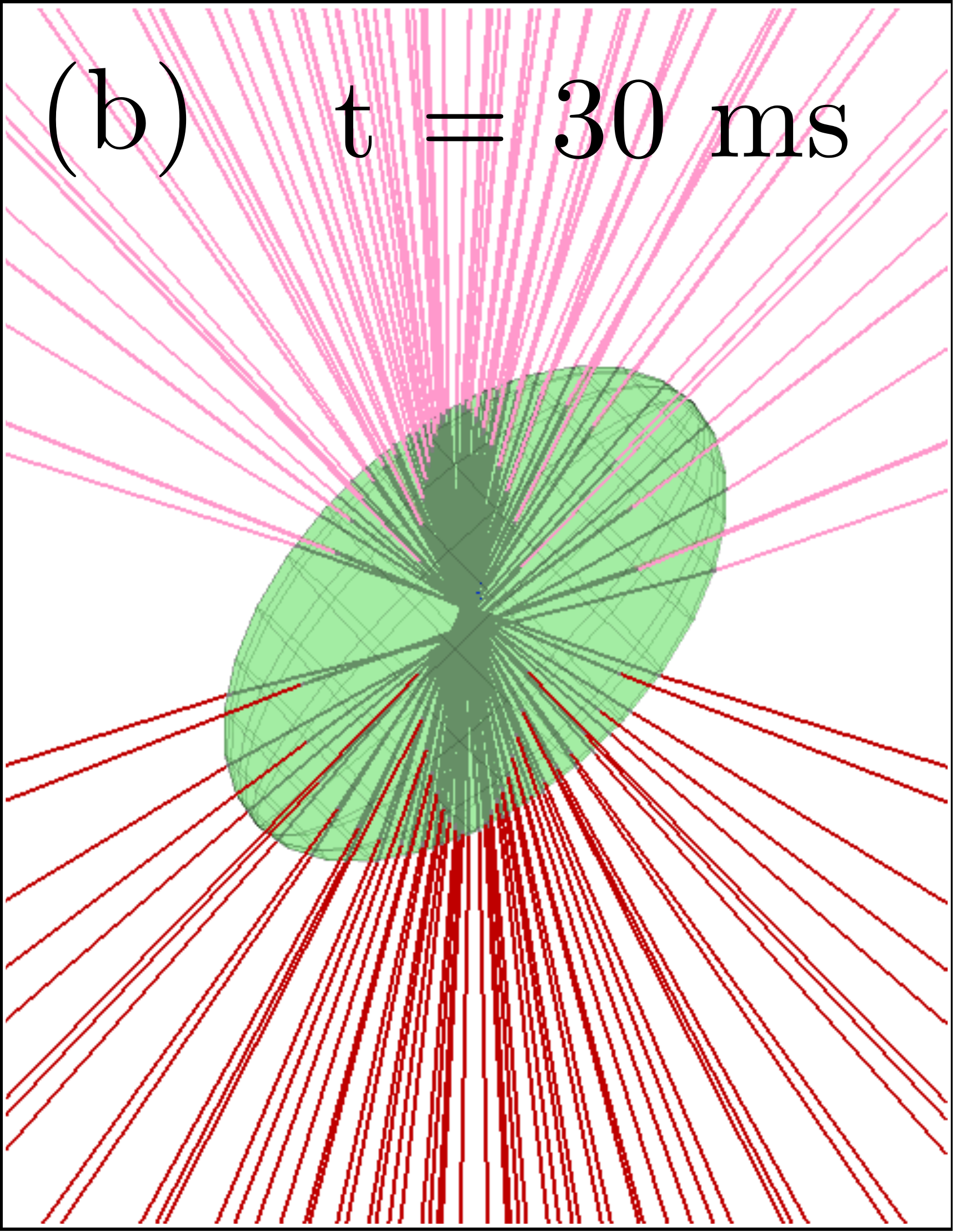}
\includegraphics[width=2.8cm]{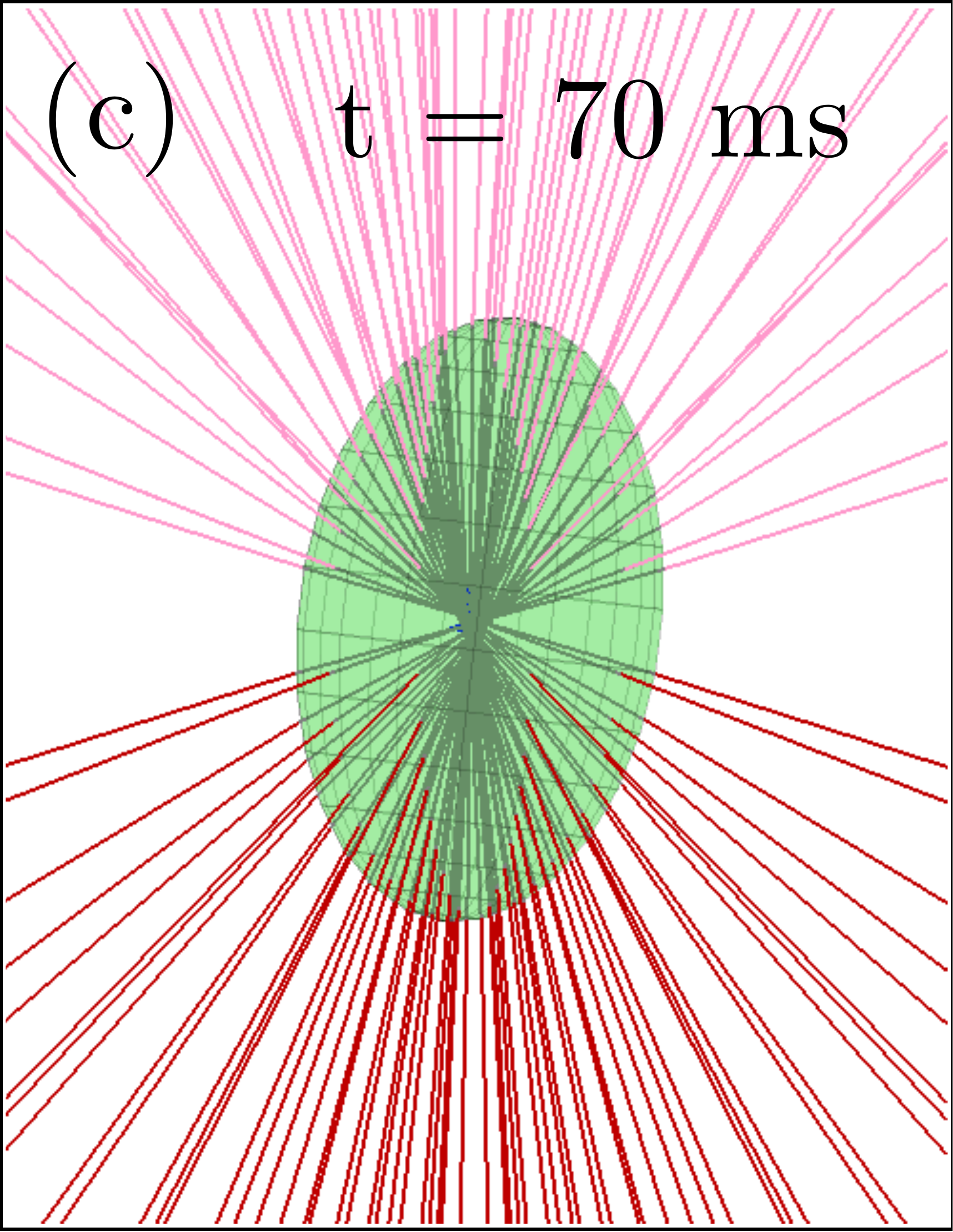}
\caption{(color online). Simulation of the motion of a prolate ellipsoidal Brownian particle ($n_{\rm p}=1.50$, short semiaxes $2.00\,{\rm \mu m}$, long semiaxis $3.33\,{\rm \mu m}$) in water ($n_{\rm m}=1.33$) under the action both of Brownian motion and of the optical forces and torques arising from a highly focused Gaussian beam (${\rm NA}=1.30$, $P_{\rm i}=1\,{\rm mW}$), whose rays are coming from the bottom. Because of the presence of optical torque, after $\sim 70\,{\rm ms}$ the particle's long axis gets aligned along the longitudinal direction. See also supplementary video 8 \cite{codes}.}
\label{fig:EllipsoidAllignment}
\end{figure}

We will first consider the case of a non-spherical particle. If the particle is convex, one can still use the formula in Eq.~(\ref{eq:DefScatGradForce}) to calculate the forces due to a single ray. However, in general, the scattered rays and the force will not lie all on incidence plane and, therefore, apart from the optical force also an optical torque can arise:
\begin{equation*} 
{\bf T} = ({\bf P}_1 - {\bf C}) \times \frac{n_{\rm m} P_{\rm i}}{c} \hat{\bf u}_{\rm i}
- ({\bf P}_1 - {\bf C}) \times \frac{n_{\rm m} P_{\rm r}^{(1)}}{c} \hat{\bf u}_{\rm r}^{(1)}  - 
\end{equation*}
\begin{equation} \label{eq:TonParticle}
\hspace*{0.2cm} -  \sum_{j=2}^{\infty} ({\bf P}_j - {\bf C}) \times \frac{n_{\rm m} P_{\rm t}^{(j)}}{c} \hat{\bf u}_{\rm t}^{(j)}
\end{equation}
where ${\bf C}$ is center of mass of the particle and ${\bf P}_{\rm j}$ is the position where the $j$-th scattering event takes place. This is different from the case of a spherical particle, such as the one shown in Fig.~\ref{fig:RayOnSphere}(a), where the torque is null \cite{Ashkin1992}. The typical order of magnitude of the torque on a particle with characteristic dimension of $\sim 1\,\mu{\rm m}$ is approximatively $10^{-18}\, {\rm Nm}$ to $10^{-21}\, {\rm Nm}$ for a ray of power $\approx 1 {\rm m}{\rm W}$, as shown in experiments \cite{Oroszi2006,Rohrbach2005,Volpe2006,Neuman2004}. For example, we can consider the case of an elongated particle, which can be modelled as a prolate ellipsoidal glass particle (short semiaxes $2.00\,{\rm \mu m}$, long semiaxis $3.33\,{\rm \mu m}$, $n_{\rm p}=1.50$, $n_{\rm m}=1.33$), as shown in Fig.~\ref{fig:EllipsoidAllignment}. Elongated particles are known to get aligned with their longer axis along the longitudinal direction because of the presence of an optical torque \cite{Borghese2008}. We simulated the optical forces using \texttt{OTGO} and the Brownian motion using the approach described in Ref.~\cite{Fernandes2002} using the freeware \texttt{HYDRO++} to calculate the diffusion tensor \cite{Hydro++}, assuming the particle to be a room temperature ($T = 300\,{\rm K}$). We start from a configuration where the particle center of mass is at the focal point, but the longer semiaxis lies in the transverse plane, as shown in Fig.~\ref{fig:EllipsoidAllignment}(a). Because of the presence of the optical torque due to a focused Gaussian laser beam of power $1\,{\rm mW}$, the particle gets aligned with the long semiaxis along the longitudinal direction in about $70\,{\rm ms}$, as shown in Figs.~\ref{fig:EllipsoidAllignment}(b) and \ref{fig:EllipsoidAllignment}(c), which is a result comparable to experiments \cite{Bayoudh2003}.

%
%
\begin{figure}[t!]  
\centering
    \includegraphics[width=8.5cm]{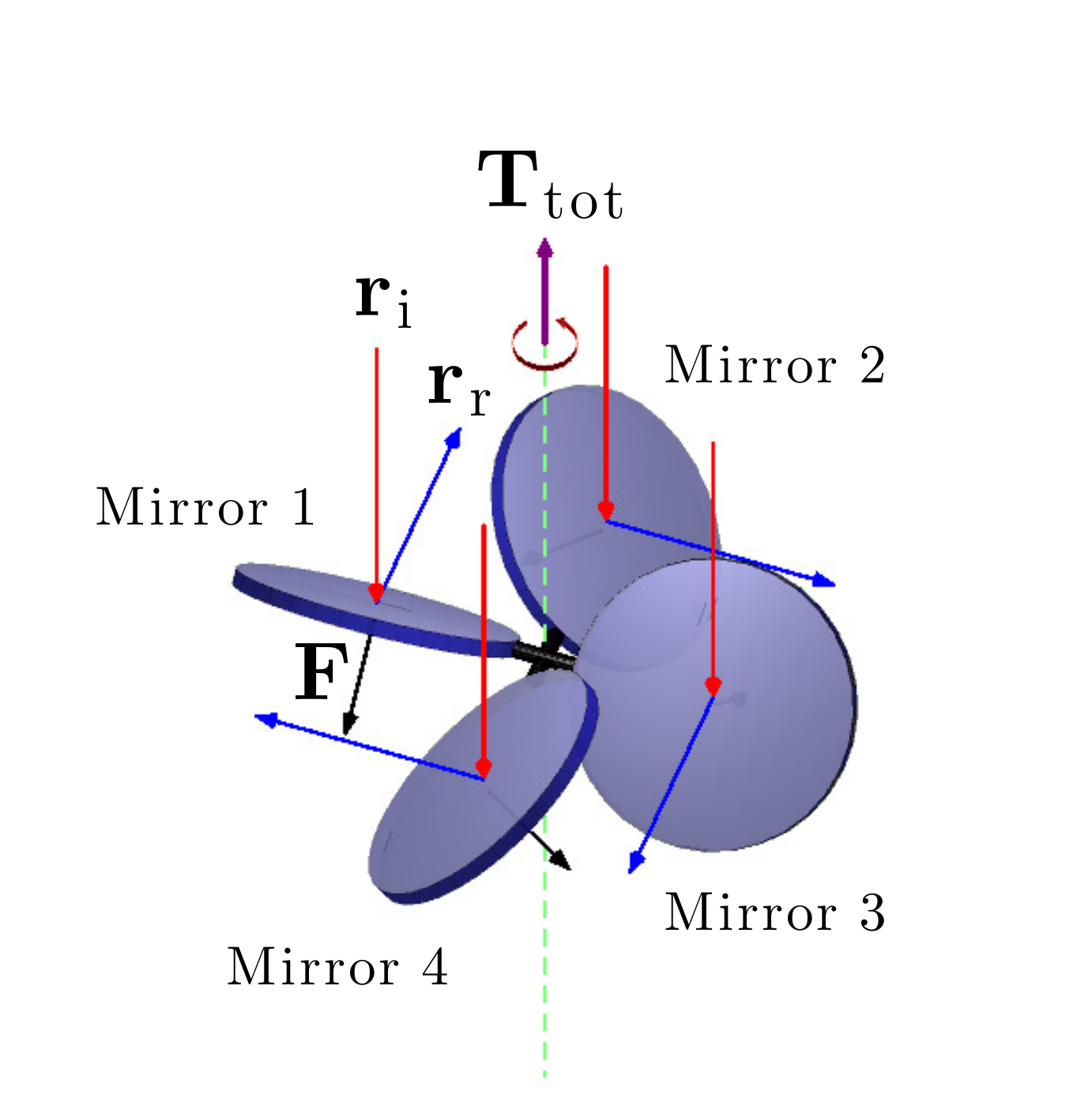}
\caption{(color online). The windmill effect. An asymmetric object illuminated by a plane wave, i.e., by a set of parallel rays (coming from the top), undergoes an optical torque that can set it into rotation. The object in the illustration is characterized by a rotational symmetry that permits the cancellation of the optical forces in the transverse plane, while an optical torque is still present.}
\label{fig:windmill}
\end{figure}

Closely related to the optical torque, the windmill effect \cite{Nieminen2006}, where an asymmetric object illuminated by a plane wave, i.e., a series of parallel rays, can start rotating around its axis, can also be reproduced using \texttt{OTGO}. In the simulation, we took a set of parallel rays and shone them onto a perfectly reflecting object reproducing the shape of a windmill wheel, i.e., four circular mirrors oriented as shown in Fig.~\ref{fig:windmill}. In the presence of an illuminating electromagnetic field, this object starts rotating. Similar structures have been indeed experimentally realized \cite{Nieminen2006,Galajda2001}.

%
%
\begin{figure}[h]  
\centering
\includegraphics[width=8.5cm]{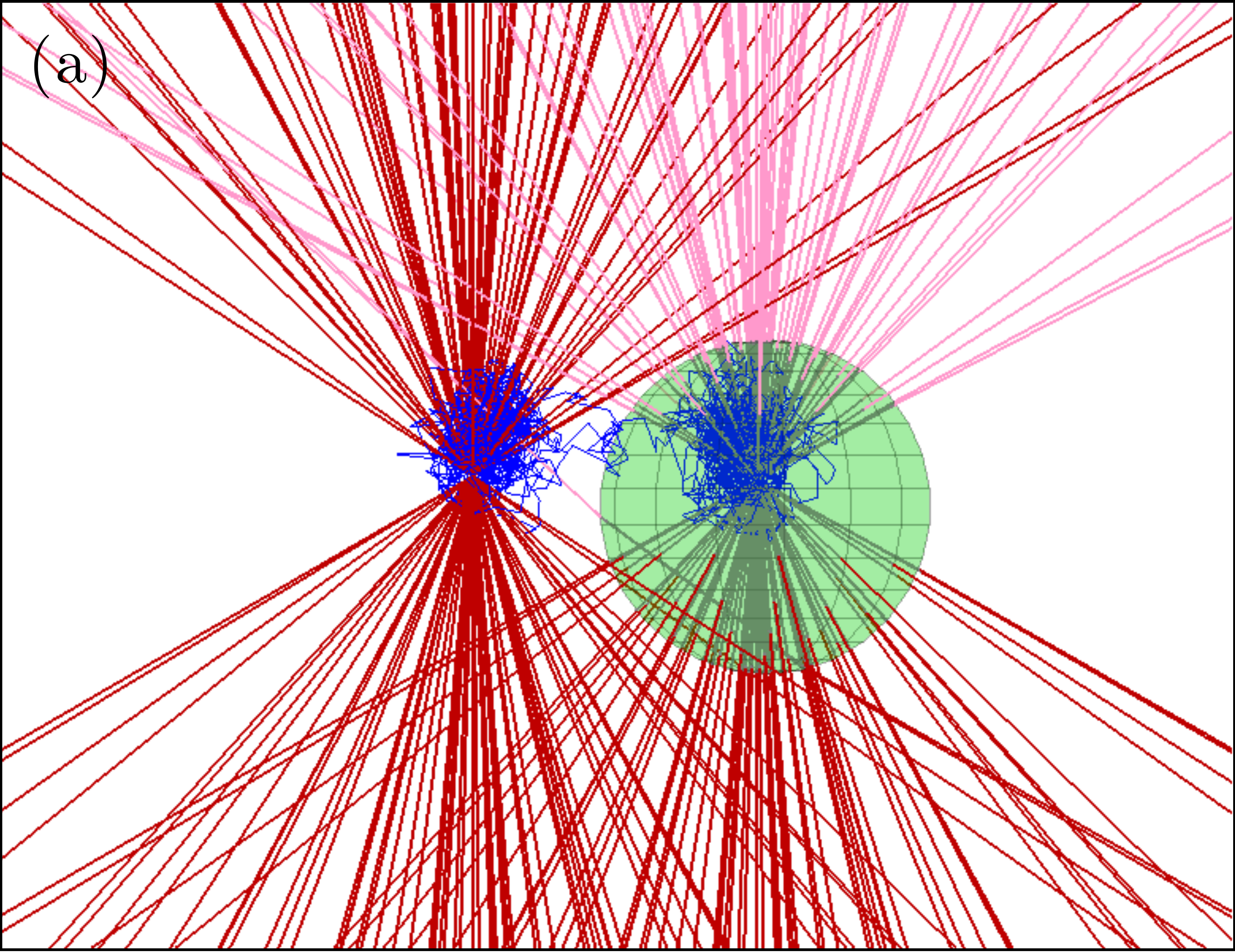}
\includegraphics[width=8.5cm]{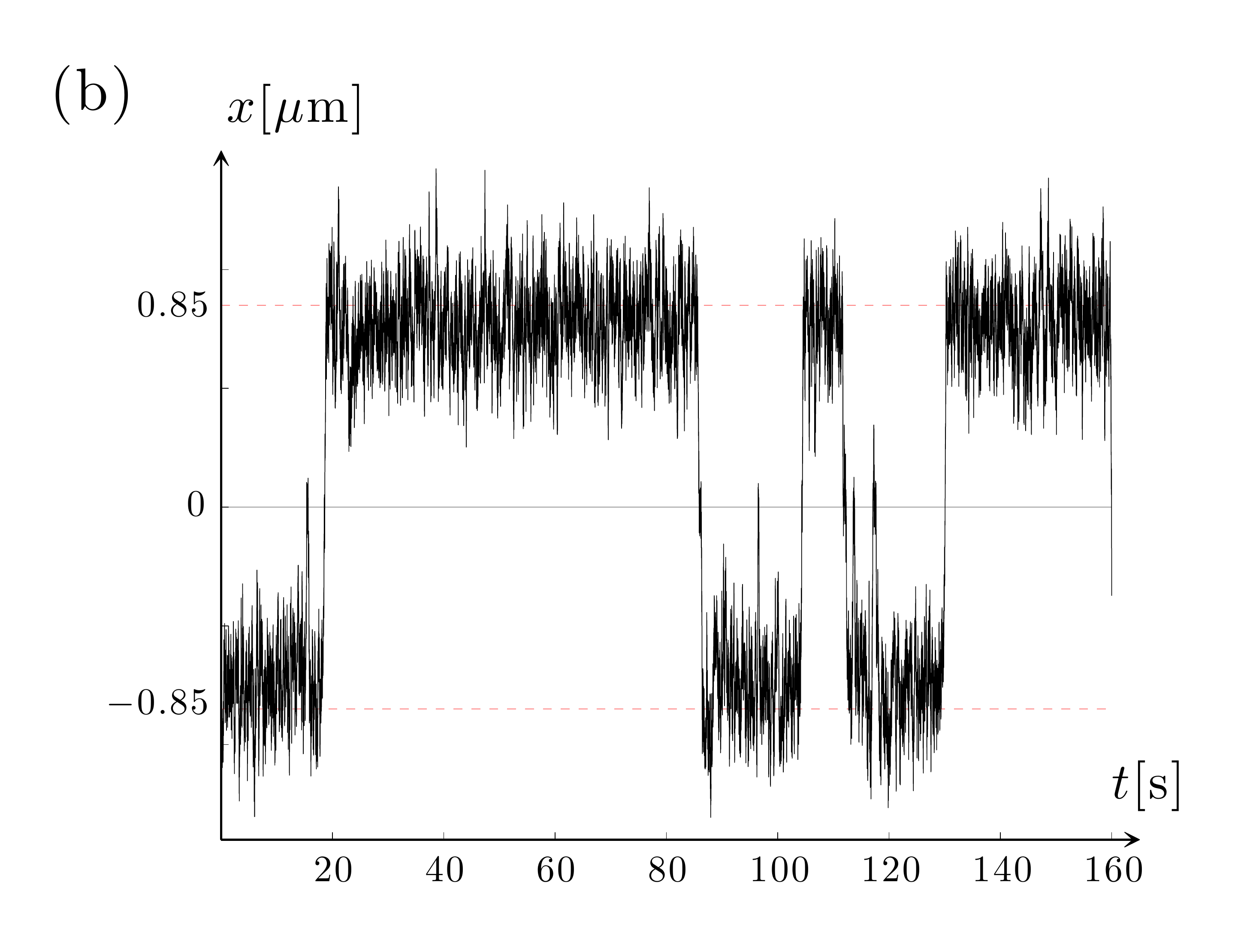}
\caption{(color online). (a) Kramers' transitions of a spherical particle of radius $R=1\,{\rm \mu m}$ ($n_{\rm p}=1.50$) in water ($n_{\rm m}=1.33$) held in the optical potential generated by a double optical tweezers, i.e., two highly focused Gaussian beams (NA=1.30, $P_{\rm i}=0.2\,{\rm m}{\rm W}$) whose focal points are separated in the transverse plane by a distance $d=1.7\,{\rm \mu m}$ and whose rays are coming from the bottom. The position of the particle is illustrated by the solid line and shows that the particle jumps between two equilibrium positions. (b) Particle position along the transverse axis that joins the two traps centers as a function of time. See also supplementary movie 8 \cite{codes}.}
\label{fig:KramersTr}
\end{figure}

Another interesting effect that can be reproduced using \texttt{OTGO} is the emergence of  Kramers' transitions \cite{Kramers1940,McCann1999}. We simulated the motion of a Brownian spherical particle with radius $R = 1\,{\rm \mu m}$ in the presence of a double trap obtained by focalizing two Gaussian beams each with power $0.25\,{\rm mW}$ so that their focal points laid at a distance $d=1.7\,{\rm \mu m}$ in the transverse plane, as shown in Fig.~\ref{fig:KramersTr}(a). Letting the system free to evolve under the action of the Brownian motion and the optical forces, the particle jumps from one potential well to the other one, as shown by the trajectory in Fig.~\ref{fig:KramersTr}(b). The relatively low value of the power of the trapping beams, necessary in order to be able to observe the transitions at room temperature $T=300\,{\rm K}$ within a relatively short time frame, is comparable with the one in actual experiments \cite{McCann1999}. Changing the parameters of the system, e.g., distance between the focal spots, beam power, temperature of the system, one can alter the transition rates and, moreover, an additional local minimum may arise between the two traps (e.g., for $d=1.5 R$), which has indeed been observed in experiments \cite{Stilgoe2011}.

%
%
\begin{figure}[t]  
\centering
\includegraphics[width=8.5cm]{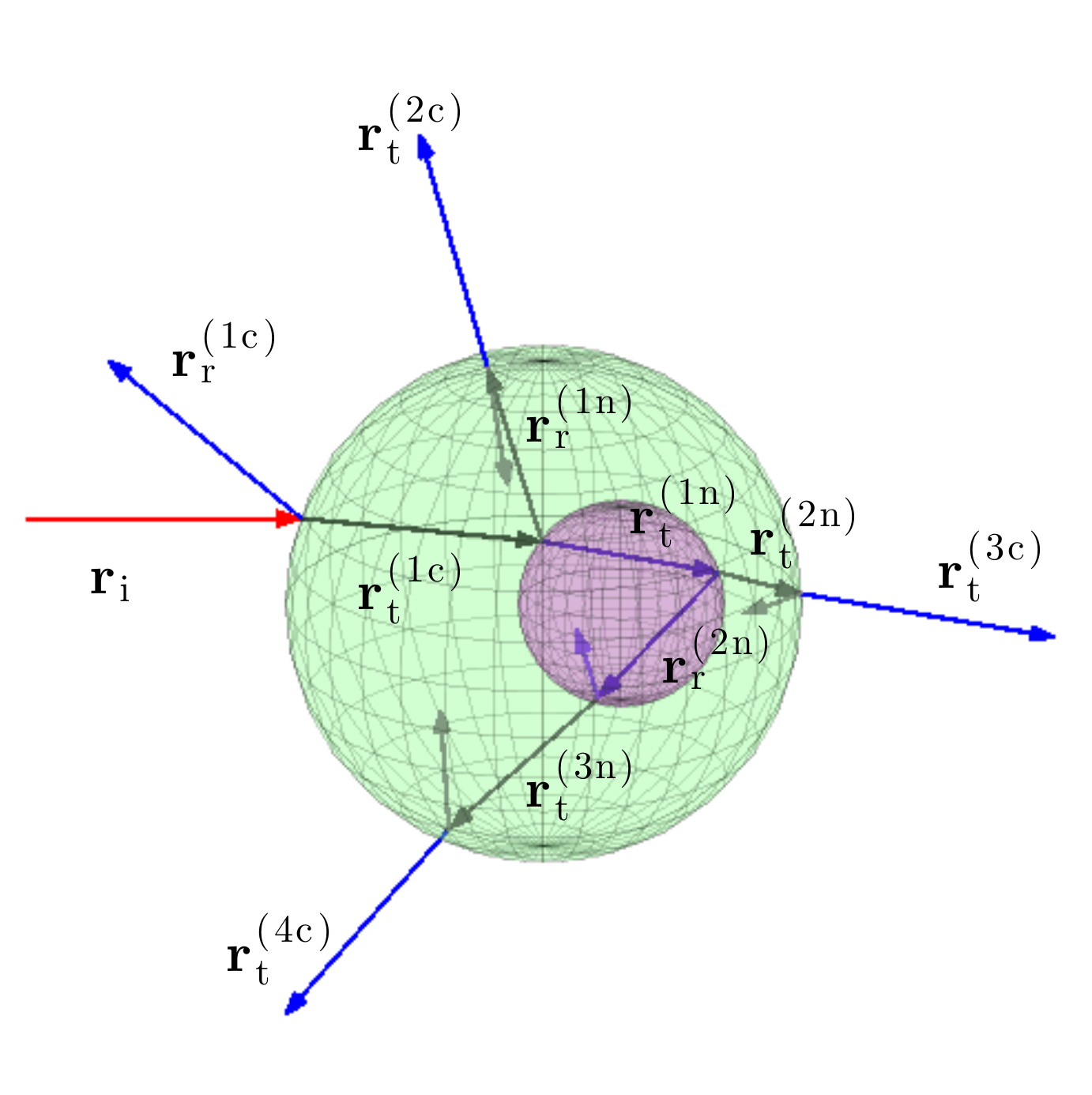}
\caption{(color online). A biological cell can be modeled by using two spherical particles with different refractive indices placed one inside the other to represent the cytoplasm and the nucleus. As the cytoplasm is a non-convex shape, the scattering process is more complex than in the case of a sphere and for a given ray multiple scattering events may need to be taken into consideration.}
\label{fig:cell}
\end{figure}

It is also possible to extend the computational capabilities of \texttt{OTGO} to non-convex and/or non-simply-connected shapes. For example, in Fig.~\ref{fig:cell} we show the case of a simple optical model for a biological cell \cite{Chang2006}: in first approximation, a cell containing a nucleus can be modeled by a sphere (the cytoplasm) containing a smaller sphere of different refractive index (the nucleus). It is interesting to notice that in a scattering event a ray can now be split into multiple rays that may not necessarily be able to escape the particle; this is typical of all non-convex shapes and can lead to a steep increase of the number of rays to be taken into account.

\subsection*{Acknowledgements}

We would like to thank Sevgin Sak{\i}c{\i} for her help in the early stages of the development of the code, and Onofrio M. Marag\'{o}, Philip H. Jones and Rosalba Saija for useful discussions and suggestions. This work has been partially financially supported by the Scientific and Technological Research Council of Turkey (TUBITAK) under Grants 111T758 and 112T235, Marie Curie Career Integration Grant (MC-CIG) under Grant PCIG11 GA-2012-321726, and COST Actions MP-1205 and IP-1208.

\bibliographystyle{unsrt}
\bibliography{GOpap_biblio}

\end{document}